\documentclass[11pt]{article}

\usepackage{amssymb}
\usepackage{subfig}
\usepackage{color}
\usepackage{epsfig,amssymb,amsfonts,amsmath,graphicx,dsfont,cite,xfrac}
\usepackage{authblk}
\definecolor{mygray}{gray}{0.5}
\usepackage{cite}
\usepackage[colorlinks=true,linkcolor=blue,citecolor=red]{hyperref}

\title{Dirac fermions in armchair graphene nanoribbons \\ trapped by electric quantum dots}

\author[${}$]{V\'i{t} Jakubsk\'y$^{1}$, \c Sengul Kuru$^2$, Javier Negro$^3$}

\affil[${1}$]{\footnotesize Nuclear Physics Institute, Czech Academy of Science, 250 68 \v{R}e\v{z}, Czech Republic}
\affil[${2}$]{\footnotesize
	 Department of Physics, Faculty of Sciences, Ankara University, 06100 Ankara, Turkey }
\affil[${3}$]{\footnotesize
		Departamento de F\'i{}sica Te\'orica, At\'omica y \'Optica, Universidad de Valladolid, 47011 
	Valladolid, Spain}

\date{}

\begin{document}
	
	\maketitle

\begin{abstract}
We study the confinement of Dirac fermions in  armchair graphene nanoribbons by means of a quantum-dot-type electrostatic potential. With the use of specific projection operators, we find exact solutions for some bound states that satisfy appropriate boundary conditions. We show that the energies of these bound states belong either to the gap of valence and conducting bands or they represent BIC's (bound states in the continuum) whose energies are embedded in the continuous spectrum. 
\end{abstract}

\section{Introduction \label{intro}}

It may be difficult to confine Dirac fermions in graphene by an electrostatic field. They can tunnel it   without being reflected.  Nevertheless, it was shown in a series of papers that the confinement is actually possible. For example, electrostatic barriers with translational symmetry can produce wave guides for strongly localized Dirac fermions \cite{portnoi1,portnoi1b,portnoi2,schulze}.  Confinement of zero-energy Dirac fermions by electrostatic quantum dots with rotational symmetry was showed analytically in \cite{portnoi3, portnoi4, lucas}. Electronic transport of Dirac fermions in armchair nanoribbons in presence of a rectangular electrostatic barrier dependent on a single coordinate was studied in \cite{zhou}. A similar setting with an additional inhomogeneity of Fermi-velocity was taken into account in \cite{nascimento}.
		
In the current letter, we present a model of Dirac fermions in armchair graphene nanoribbons (AGNR) under a localized electrostatic field. 
Although it lacks both translational and rotational symmetry, we show  that this system can confine bound states with non-zero energies that are either in the spectral gap or belong to the continuum spectral band and correspond to BIC (bound states in the continuum). The corresponding wave functions are also given by analytical closed expressions.
In order to facilitate their calculation, we propose the method of projection operators based on symmetries of the two-valley Dirac equation. The projectors map the generic solutions of the stationary equation into those that comply with the boundary conditions. This approach is rather general and can be applied to a wider class of systems then those presented in this work. 

The stationary equation of free Dirac massless fermions in graphene can be written in the following form \cite{Akhmerov}: 
\begin{equation}\label{planestac}
H_0\Psi=(\tau_0\otimes\sigma_1 p_x+\tau_0\otimes\sigma_2 p_y)\Psi=E\Psi \,,
\end{equation}
where both $\tau_j$ and $\sigma_j$, $j=1,2,3$, are Pauli matrices, $\tau_0$ and $\sigma_0$ are $2\times 2$ identity matrices. The matrices $\tau_j$ act on the valley degree of freedom associated to the Dirac points ${\bf K}$ and ${\bf K}'$ of graphene. The matrices $\sigma_j$ act on the sublattice degree of freedom related to presence of two triangular sublattices ${\bf A}$ and ${\bf B}$. For convenience, we identify the components $\psi{(')}_{X}$ of $\Psi$ in (\ref{planestac}) with the sublattices and valleys as follows, see \cite{Akhmerov}:
\begin{equation}\label{psiB}
\Psi:=(\psi_A,\psi_B,\psi'_B,-\psi'_A)\,.
\end{equation}
For instance, $\psi_A(')({\bf r})$ describes Dirac fermions localized at the lattice $A$ with momentum near the Dirac point ${\bf K}(')$. The interpretation of the other components goes in the same manner.

In graphene nanoribbons, the solutions of (\ref{planestac}) are subject to the boundary  conditions that characterize the edges of the graphene strip. The boundary conditions were discussed in a number of works \cite{ Akhmerov,Brey,siegl,Falco}.	We adopt here the notation of \cite{Akhmerov} in which the stationary equation of the free Dirac fermion coincides with (\ref{planestac}) \footnote{
	Hamiltonian used by Brey et al in \cite{Brey} reads as
	$H_{B}=\tau_3\otimes \sigma_1 p_x+\tau_0\otimes \sigma_2 p_y$
	whereas the Hamiltonian used by Akhmerov et al in \cite{Akhmerov} is
	$H_{A}=\tau_0\otimes (\sigma_1p_1+\sigma_2p_2)$.
	The mapping between the two operators is mediated by the operator
	$U=\left(\begin{array}{cc}\sigma_0&0\\0&-\sigma_2\end{array}\right).$
	There holds
	$H_{B}=UH_{A}U^{-1}.$
}.

The boundary condition at the boundary $\Gamma$ can be specified in terms of a unitary matrix $M$ as follows:
\begin{equation}\label{BC}
\Psi|_{\vec{x}\in\Gamma}=M\Psi|_{\vec{x}\in\Gamma},\quad M^{\dagger}=M,\quad M^2=1.
\end{equation}
In order to avoid the leaks of probablity current through the boundary, the matrix has to satisfy 
\begin{equation}
\{M,\mathbf{n_B}.\mathbf{J}\}=0,\label{Mn_bJ}
\end{equation}
where $\mathbf{n_B}=\{n_1,n_2,0\}$ is the normal vector to the boundary and  $\mathbf{J}=(\tau_0\otimes\sigma_1,\tau_0\otimes\sigma_2,0)$ is the current density operator. Together with the requirement of the time-reversal symmetry \cite{Akhmerov}, the matrix $M$ acquires the following general form
\begin{equation}\label{MTinv}
M=\boldsymbol{\nu}\cdot\boldsymbol{\tau}\otimes \mathbf{n_1}\cdot\boldsymbol{\sigma},
\end{equation}
where $\boldsymbol{\tau}=\{\tau_1,\tau_2,\tau_3\}$, $\boldsymbol{\sigma}=\{\sigma_1,\sigma_2,\sigma_3\}$; $\mathbf{n_1}$ is a unit vector perpendicular to $\mathbf{n_B}$ and $\boldsymbol{\nu}$ is an arbitrary three-dimensional unit vector. Let us mention two specific families of the boundary conditions. In the first case, the boundary condition does not mix the valleys,
\begin{equation}\label{zig-zag-like M}
M=\tau_3\otimes \sigma_3 e^{i\nu \sigma_1},\quad \nu\in(-\pi,\pi].
\end{equation}
When $\nu=0$, $M$ represents zig-zag boundary condition. For $\nu=\pm\pi/2$, it describes infinite-mass (so called MIT) boundary condition. The other values of $\nu$ can be attributed to the presence of staggered potential at the boundary, see \cite{Akhmerov}.
The second family corresponds to armchair-like boundary,
\begin{align}\label{generalM}
M&=
\tau_1e^{i\alpha \tau_3}\otimes \sigma_1 e^{i\mu \sigma_3},\quad \alpha,\mu\in(-\pi,\pi].
\end{align}   
Here $\mu$ depends on the explicit form of $\mathbf{n_B}$ whereas the parameter $\alpha$ can be attributed to the width $L$ of the nanoribbon.

It is worth mentioning that the armchair boundary conditions are less universal than zig-zag graphene nanoribbons (ZGNR)  boundary  conditions with $M=\tau_3\otimes\sigma_3$, which apply to a wider range of lattice terminations \cite{Akhmerov}. Nevertheless, AGNRs  attract attention due to the energy gap that opens between positive and negative energies, which is desirable in electronic devices.  Additionally, it proved to be possible to fabricate highly precise AGNR in the experiments  \cite{experimentalAGNR}.

We will present a model where Dirac fermions in AGNR are confined by strongly localized electrostatic field. But first, it is convenient to review briefly the spectral properties of free Dirac fermions in an armchair nanoribbon, since it will be quite helpful for the following analysis of the electric quantum dots.

\section{Free particle in AGNR}
Let us consider an armchair nanoribbon of  width $L$ which is oriented along $y$-axis, i.e. $x\in[0,L]$ and $y\in\mathbb{R}$. The normal vector $\mathbf{n_B}$ perpendicular to the boundary and pointing outwards is $\mathbf{n_B}=(-1,0,0)$ for $x=0$ and $\mathbf{n_B}=(1,0,0)$ for $x=L$. Therefore, we can see from (\ref{Mn_bJ}) and (\ref{generalM}) that the matrices $M_1$ and $M_2$ fixing the boundary conditions at $x=0$ and $x=L$, respectively, can be defined without lack of generality as\footnote{Indeed, if the wave function $\Psi$ satisfies $(1-M_1)\Psi(0,y)=0$ and $(1-M_2)\Psi(L,y)=0$, then the wave function $\tilde{\Psi}=e^{i\beta \tau_3}\Psi$ satisfies $(1-\tilde{M}_1)\tilde{\Psi}(0,y)=0$ and $(1-\tilde{M}_2)\tilde{\Psi}(L,y)=0$ where $\tilde{M}_{1}=\tau_1e^{-2i\beta\tau_3}\otimes \sigma_2$ and $\tilde{M}_{2}=\tau_1e^{i(\alpha-2\beta)\tau_3}\otimes \sigma_2$.}
\begin{equation}\label{M1M2}
M_1=\tau_1\otimes\sigma_2,\quad M_2=\tau_1e^{i\alpha \tau_3}\otimes\sigma_2,\quad \alpha\in\left(-\pi,\pi \right].
\end{equation}
For the armchair nanoribbon without any additional interaction at the boundary, the parameter $\alpha$ can be associated with the width $L$ of the nanoribbon. Its origin can be traced back to the tight-binding model. Indeed, the solutions  of the Dirac equation (\ref{planestac})  correspond to the slowly varying amplitudes $\psi_A{(')},\psi_B{(')}$ of the total wave functions defined on the sublattices $A$ and $B$, see \cite{Brey, NovoselovGuinea},
\begin{equation}\Psi_A=e^{i\, K x}\psi_A-ie^{i\, K' x}\psi_A',\quad\Psi_B=e^{i\, K x}\psi_B-ie^{i\, K' x}\psi_B'.\end{equation} 
Remember that the Dirac points are $
{\bf K}= (K,0)=(\frac{2\pi}{3 a_0},0)$ and ${\bf K}'= (K',0)=-{\bf K}$
where $a_0$ is the constant of the lattice.
The armchair boundaries are formed by the atoms from  both the sublattices $A$ and $B$. The corresponding total wave functions should be vanishing there, 
\begin{equation}\label{bf}
\left(\psi_X(x,y) e^{i K x}-i \psi'_X(x,y) e^{i K' x}\right)\vert_{{}^{x=0}_{x=L}} =0,\quad X = A, B,\  
\forall y\,.
\end{equation} 
Comparing these relations with (\ref{M1M2}), we get $\alpha=2KL=\frac{4\pi}{3a_0}L$. Therefore, for  nanoribbons with no additional edge interactions, we get effectively $\alpha=0$ for $L=3Ma_0$ and $\alpha=\pm 2\pi/3$ for $L=(3M\pm 1)a_0$ where $M$ is an integer, see \cite{Brey, NovoselovGuinea}. Other values of $\alpha$ can reflect the presence of an additional interaction  or altered crystal structure at the boundary \cite{vanOstaay}. 

The admissible solutions of the stationary equation (\ref{planestac}) have to satisfy 
\begin{equation}
H_0\Phi=E\Phi,\quad \Phi(0,y)=M_1\Phi(0,y),\quad \Phi(L,y)=M_2\Phi(L,y) \ .
\end{equation}
The boundary conditions (10) have been replaced by matrix equations which allow for an easier algebraic manipulation, as it will be shown in the following. This matrix form will play a fundamental role in our approach.
We can find the wave functions $\Phi(x,y)$ in the following manner:
the Hamiltonian $H_0$ commutes with the operator $P_0M_1$, $[H_0, P_0M_1]=0$,
where $P_0$ is reflection along the axis $y=0$, i.e. $P_0f(x,y)=f(-x,y)$. We can utilize this fact and define the wave function
\begin{equation}\label{Phi}
\Phi(x,y)=(1+P_0M_1)F(x,y),
\end{equation}
where $F=F(x,y)$ is a generic solution of $(H_0-E)F(x,y)=0$. By construction, $\Phi$ solves the same equation  and additionally it satisfies the boundary condition at $x=0$. Indeed, we have
\begin{equation}
\Phi(0,y)=(1+P_0M_1)F(x,y)\vert_{x=0}=(1+M_1)F(0,y)=M_1(1+M_1)F(0,y)=M_1\Phi(0,y),
\end{equation}
where we used the fact that $M_1^2=\tau_0\otimes\sigma_0$. Hence, the operator $1+P_0M_1$ works as a projector to the space of functions where the boundary condition at $x=0$ is satisfied \footnote{At the boundary $x=0$, the space of solutions of $(H_0-E)F(x,y)\vert_{x=0}=0$ is four-dimensional. The projector in (\ref{Phi}) reduces into $1+M_1$ at the boundary and projects the space of solutions into a two-dimensional subspace.}. If we substitute the explicit form of $\Phi$ into the boundary condition at $x=L$, $(1-M_2)\Phi(L,y)=0$, we get the following equation
\begin{equation}\label{F1}
(1-M_2)(F(L,y)-M_2M_1F(-L,y))=0.\end{equation} 
Therefore, when $F(x,y)$ satisfies 
\begin{equation}\label{F(L,y)}
F(L,y)=M_2M_1F(-L,y)=e^{i\tau_3\alpha}\otimes\sigma_0\,F(-L,y),
\end{equation}
then $\Phi(x,y)$ defined by (\ref{Phi}) fulfil  the boundary conditions both at $x=0$ and $x=L$. 

Relation (\ref{F(L,y)}) requires the components of $F(x,y)$ to be quasi-periodic. 
It is rather straightforward to find the bispinors $F(x,y)$ such that (\ref{F1}) is satisfied.  They are
\begin{equation}
F_{1,n}(x,y)=e^{ik_y y+i\left(\frac{n\pi}{L}+\frac{\alpha}{2L}\right)x}\left(\frac{\frac{n\pi}{L}+\frac{\alpha}{2L}-ik_y}{\sqrt{\left(\frac{n\pi}{L}+\frac{\alpha}{2L}\right)^2+k_y^2}},1,0,0\right),\quad F_{2,n}=\tau_{1}\otimes\sigma_0\,F_{1,n}\vert_{{}^{\alpha\rightarrow-\alpha}_{n\rightarrow-n}},\quad n\in\mathbb{Z},
\end{equation}
where $
\left(H_0-E_n\right)F_{1(2),n}=0$ and $E_n=\sqrt{\left(\frac{n\pi}{L}+\frac{\alpha}{2L}\right)^2+k_y^2}.$
The corresponding bispinors that comply with the boundary conditions are 
\begin{equation}
\Phi_{1,n}=(1+P_0M_1)F_{1,n}\,,\quad \Phi_{2,n}=(1+P_0M_1)F_{2,n},
\end{equation}
such that $(H_0-E_n)\Phi_{j,n}=0$, $j=1,2$.

The spectrum of a generic self-adjoint operator consists of two disjoint sets; the essential spectrum $\sigma_{\rm ess}$ and the discrete spectrum $\sigma_d$. In case of the free-particle Hamiltonian $H_0$ with the domain specified by the boundary conditions (\ref{M1M2}), there are no discrete eigenvalues. Its spectrum is formed just by the essential spectrum. The gap between positive and negative energies depends on the value of the parameter $\alpha$, 
\begin{equation}\label{sigmaess}
\sigma(H_0)=\sigma_{\rm ess}(H_0)=
(-\infty,-E_0]\cup[E_0,\infty),\quad E_0=\frac{|\alpha|}{2L}.
\end{equation}
When $\alpha=0$, i.e. $M_2=M_1$ there is no gap in the spectrum so that the nanoribbon is metallic. When $\alpha=\pi$, i.e. $M_2=-M_1$, the spectral gap is maximal.  

It is worth noticing that the construction of eigenstates via the projector (\ref{Phi}) is rather general and can be applied to a large class of energy operators, see Appendix. Formula (\ref{F(L,y)}) suggests that it can be particularly useful for periodic systems where the wave functions are quasi-periodic due to the Bloch theorem. We will apply this construction in the next section.


\section{Electrostatic quantum dots in AGNR}
Now, let us analyze the possible confinement of Dirac fermions on armchair nanoribbons  in the presence of electrostatic quantum dots described by the following stationary equation,
\begin{equation}\label{Hsuper}
H\Psi=(H_0+V(x,y))\Psi=E\Psi,
\end{equation}
where $H_0$ is the free Hamiltonian given in (\ref{planestac}) 
and the potential term has the following explicit form,
\begin{equation}\label{V(x,y)}
V(x,y)=-\sigma_0\otimes\sigma_0\frac{4m\omega^2\sin^2\kappa x }{m^2+\omega^2\cos 2\kappa x+\kappa^2\cosh 2\omega y},\quad \kappa=\sqrt{m^2+\omega^2}\,.
\end{equation}
The parameters $m$ and $\omega$ can acquire arbitrary nonvanishing real values.
The Hamiltonian commutes both with $P_0M_1$ and  $P_0M_2$,
\begin{equation}
[H,P_0M_1]=[H,P_0M_2]=0,
\end{equation}
where $M_1$ and $M_2$ are given in (\ref{M1M2}).
The potential term corresponds to an electrostatic potential which is periodic in $x$ and
decreases very fast in $y$,
\begin{equation}\label{t}
V(x,y)=V(x+T,y),\quad T=\frac{\pi}{\kappa},\qquad\quad
V(x,y) \ \to 0, \ {\rm for}\ |y| \to \infty\,.
\end{equation}
As the potential term is exponentially vanishing in $y$, its presence does not alter the essential spectrum of Dirac fermions.  Therefore, there holds
\begin{equation}
\sigma_{\rm ess}(H)=\sigma_{\rm ess} (H_0),
\end{equation} 
see Lemma 3.4 in \cite{siegl2}. However, there can emerge bound states with energies either in the gap or within the essential spectrum due to the interaction. 

It was shown in \cite{superKlein} that the stationary equation (\ref{Hsuper}) can be solved for $E=m$. Besides scattering states, there are also four solutions that are strongly localized by the electrostatic field. The first one, $v_1$, has the following expression,
\begin{eqnarray}
&&v_1(x,y)=\frac{1}{D(x,y)}\left(\begin{array}{c}-i\cosh(\omega y)[\kappa\cos (\kappa x)+m\sin (\kappa x)]+\omega \sin (\kappa x)\sinh (\omega y)
\\[1.ex]
\cosh (\omega y)[-\kappa\cos (\kappa x)+m\sin (\kappa x)]+i\omega \sin( \kappa x)\sinh (\omega y)
\\[1.ex]
0
\\[1.ex]
0\end{array}\right),\label{v1}
\end{eqnarray}
where we abbreviated the denominator by the nonvanishing function $D(x,y)=m^2+\omega^2\cos (2\kappa x)+\kappa^2\cosh (2\omega y)$. The other solutions, $v_2,v_3$ and $v_4$
are obtained from $v_1$ by
\begin{eqnarray}
&&v_2(x,y)=\sigma_0\otimes\sigma_3v_1(-x,-y),\quad v_3=\sigma_1\otimes\sigma_0v_1,\quad v_4=\sigma_1\otimes\sigma_0v_2.\label{v_a}
\end{eqnarray}
All of them have the same energy $m$,
\begin{equation}\label{superKleinstac}
Hv_a(x,y)=m\,v_a(x,y),\quad a=1,2,3,4.
\end{equation}
These wave functions are strongly localized along the $y$-axis and are $T$-anti-periodic in the $x$-direction,
\begin{equation}\label{v_aantiper}
v_{a}(x,y)=-v_a(x+T,y).
\end{equation}
Let us stress that states $v_a(x,y)$ do not necessarily comply with the boundary conditions of the armchair nanoribbon. Nevertheless, we will consider two specific cases where localized states based on (\ref{v1}) and (\ref{v_a}) can be found.

\subsection{AGNR with the maximal band gap}

\begin{figure}[h!]
	\begin{center}
		\includegraphics[scale=0.4]{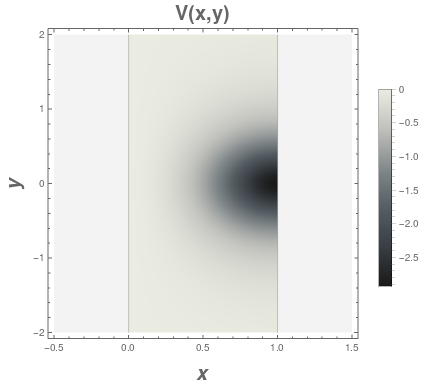} 
		\ \qquad \includegraphics[scale=0.4]{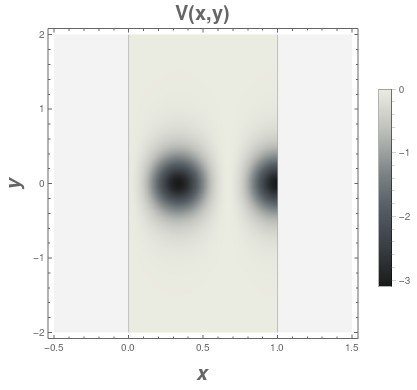}
		\\
		\includegraphics[scale=0.4]{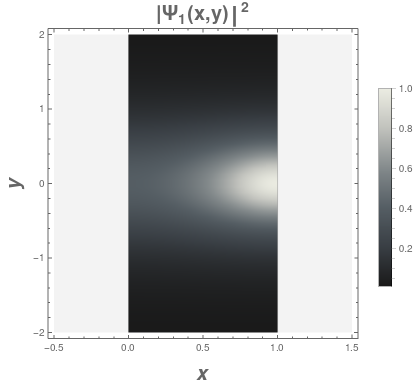} 
		\qquad \qquad \includegraphics[scale=0.4]{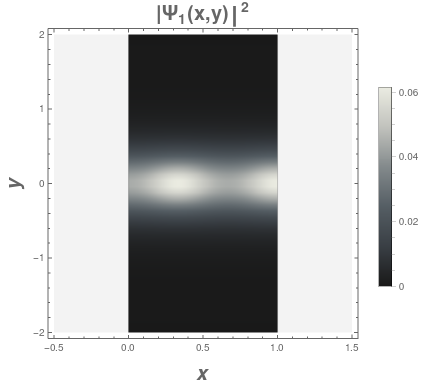}
	\caption{ Up: Density of probability of (not normalized) $\Psi_1$ in (\ref{Psi}) for the boundary matrices $M_2=-M_1$. On the left, the potential has period $T_0=2L$. The corresponding energy level $E\equiv m$ belongs to the discrete spectrum.   On the right, there holds $T_1=\frac{2L}{3}$. We fixed $m=4$ (right). In this case, the energy of the bound states  is embedded in the continuum. In both cases, $L=\frac{\pi}{2\sqrt{2}}$ and $\omega=\sqrt{\frac{\pi^2}{T^2}-m^2}$.\newline	
	Down: Electrostatic potential $V(x,y)$ in (\ref{V(x,y)}) for the aforementioned values of the parameters $\omega$, $m$ and the width $L$. 
	} \label{fig1}
\end{center}
\end{figure}

Let us focus on the specific case of the armchair-like boundary conditions (\ref{M1M2}) where the spectral gap is maximal, i.e. $\alpha=\pi$ and $M_2=-M_1$. We fine-tune the electrostatic field such that its period $T_j$ can match the fixed width $L$ of the nanoribbon through different choices, depending on a positive integer $j$, in the following manner,
\begin{equation}\label{width}
T_j=\frac{\pi}{\sqrt{m_j^2+\omega_j^2}}=\frac{L}{j+1/2},\quad j\in\{0,1,2,\dots\}.
\end{equation}
Now, we can follow the same steps as in case of the free particle model. We compose the wave functions $\Psi_a$ as 
\begin{equation}\label{Psi}
\Psi_a=(1+P_0M_1)v_a(x,y),\quad H\Psi_a=m_j\Psi_a, \quad a=1,2,3,4,
\end{equation}
which, by definition, comply with the boundary condition at $x=0$.
The eigenstates (\ref{v_a}) are $T-$anti-periodic, $v_a(x,y)=-v_a(x+T,y)$, and satisfy
\begin{equation}\label{antiper1}
v_a(L,y)=M_1M_2v_a(-L,y)=-v_a(-L,y).
\end{equation}
Then it is possible to show (see the Appendix) that $\Psi_a$
satisfies the boundary conditions,
\begin{equation}
(1-M_1)\Psi_a(0,y)=0,\quad (1+M_1)\Psi_a(L,y)=0,\quad a=1,2,3,4.
\end{equation}
Additionally, the bispinors $\Psi_a$ are square integrable on $(x,y)\in[0,L]\times\mathbb{R}$ so that they represent  bound states confined by the electrostatic field. As their density of probability is essentially indistinguishable in the plots, we illustrate just one of them in Fig.~\ref{fig1} for different choices of parameters $m,$ $\omega$ and $L$.  

The energy $E=m_j$ of the bound states (\ref{Psi}) can be either in the energy gap or it can be within the spectral band, i.e. immersed in the essential spectrum of $H$. In the later case, the bound states would form BICs of the system. 
The condition (\ref{width}) does not fix the parameters $m_j$ and $\omega_j$ uniquely. Indeed, with $j$ being fixed, any of the configurations that satisfies $ m_j^2 + \omega_j^2=\frac{\pi^2(2j+1)^2}{4L^2}$ is admissible. These configurations differ by the energy $m_j$ of the bound states $\Psi_a$,
\begin{equation} \label{m_j}
m_j \in \left(0,\frac{\pi(2j+1)}{4L}\right).
\end{equation}
The threshold of the essential spectrum  $E_0=\frac{\pi}{2L}$ stays within this interval for any $j\in\{1,2,\dots\}$, and, therefore, there exists such $m_j$ in (\ref{m_j}) such that $m_j<E_0$ or such that $m_j>E_0$ for nonzero $j$.  For $j=0$, all the admissible values of $m_0$ lie below the threshold of the essential spectrum, $m_0<E_0.$  
We arrive to the following conclusions:
\begin{itemize}
\item If the period of the potential is $T_0=2L$, then the energy of the bound state is in the interval 
$0 < m_0 < E_0=\frac{\pi}{2L}$ and, therefore, it belongs to the discrete spectrum $\sigma_d(H)$.
\item  If the period $T_j$ of the potential is smaller,  $j=1,2,\dots$, then we have two options how to choose the parameter $m_j$:
\[
\begin{array}{lc}
i) \quad  & 0 < m_j < E_0=\frac{\pi}{2L}
\\[1.ex]
ii) \quad  & E_0=\frac{\pi}{2L} < m_j < \frac{\pi(2j+1)}{2L}
\end{array}
\]
If $m_j$ satisfies i), then the energy $E_{\rm  } = m_j$ stays in the gap and belongs into $\sigma_d(H)$. When $m_j$ satisfies ii), the energy $m_j$ belongs to the essential spectrum $\sigma_{\rm ess}(H)$ and the corresponding states $\Psi_a$ represent BICs. 
\end{itemize}
We illustrate two configurations of the electrostatic dots in the nanoribbon in the Fig.\ref{fig1} where the bound state energy $E=m$ will be a discrete energy in the gap or it will be embedded into the continuum spectrum in the second case, for $j=0$ and $j=1$, respectively.

\subsection{Quantum dots in  metallic AGNR}
Let us consider now the setting where we fix $M_2=M_1$, i.e. $\alpha=0$. In this case, there is no gap in the essential spectrum (\ref{sigmaess}) and the nanoribbon is metallic.  
We alter the electrostatic field (\ref{V(x,y)}) in order to match its period with the width of the nanoribbon in the following manner,
\begin{equation}\label{width2}
T=\frac{L}{j},\quad j=1,2,3,\dots .
\end{equation}
We can achieve it by fixing $m\equiv m_j$ and  $\omega\equiv \omega_j$ such that 
\begin{equation}\label{mj2}
m_j^2 + \omega_j^2=\frac{\pi^2j^2}{L^2},\quad j=1,2,3,\dots.
\end{equation}  

\begin{figure}[h!]
	\begin{center}
\includegraphics[scale=0.45]{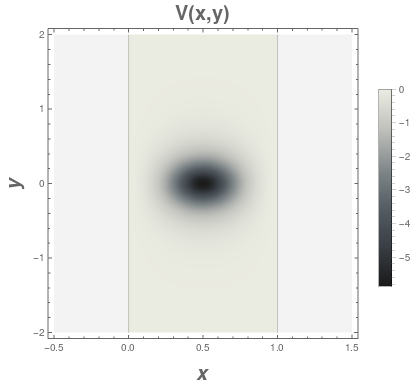} \qquad \includegraphics[scale=0.45]{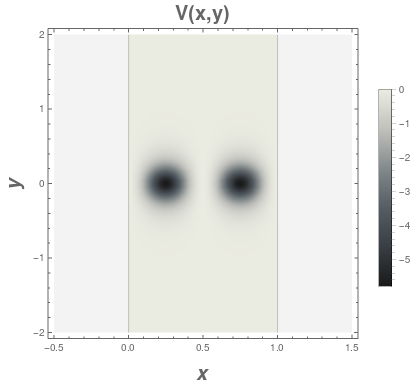}
\\
	\includegraphics[scale=0.45]{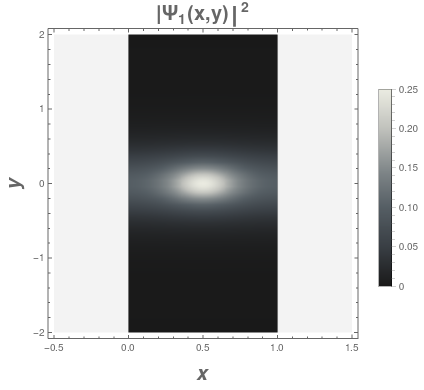} \qquad \includegraphics[scale=0.45]{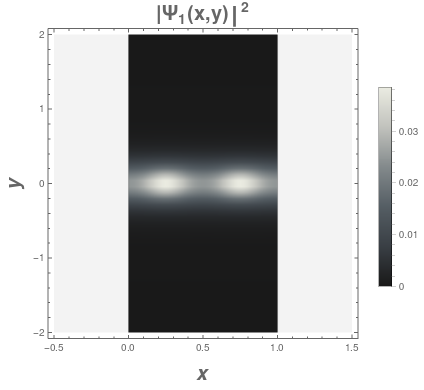}
			\caption{ Up: Density of probability of (non-normalized) $\Psi_1$ in (\ref{Psib}) for the boundary matrices $M_1=M_2$. Down: electrostatic potential $V(x,y)$ in (\ref{V(x,y)}). We fixed $m_1=2$, $T_1=L$ (left) $m_2=5$, $T_2=\frac{L}{2}$ (right). $L=1$ and  $\omega_j=\sqrt{\frac{\pi^2}{T_j^2}-m_j^2}$, $j=1,2$ in all cases.} \label{fig2}
	\end{center}
\end{figure}

In the construction of bound states confined by these quantum dots, we can follow exactly the same steps as in the previous subsection.  
We define
\begin{equation}\label{Psib}
\Psi_a=(1+P_0M_1)v_a(x,y),\quad H\Psi_a=m_j\Psi_a, \quad a=1,2,3,4.
\end{equation}
Due to (\ref{width2}), the bispinors (\ref{v_a}) are $2L$ periodic, and there holds in particular
\begin{equation}\label{antiper1}
v_a(L,y)=M_1M_2v_a(-L,y)=v_a(-L,y).
\end{equation}
It is possible to show that the the eigenstates $\Psi_a$ comply with the required boundary conditions and represent Dirac fermions confined by the electrostatic field. They are BICs as their energy $E=m_j$ is embedded in the essential spectrum,
\begin{equation}
m_j\in\sigma_{\rm ess}(H).
\end{equation}
We illustrate the potential together with the probability of density of the states in  Fig.~\ref{fig2}.

\section{Conclusions}

Our results show that the localized electric field can confine Dirac fermions with energies that are either in the gap or are embedded in the continuum.  
In the presented models, the match between the period of the electric field and the width of the nanoribbons,  (\ref{width})  or (\ref{width2}), allowed us to find the localized solutions analytically at the fixed energy $E=m$.  When the width of the nanoribbon is mismatched with the period of the potential, we expect the confined states to keep existing in the system. Nevertheless, their energy would depart from the value $m$ and their form would not be calculable analytically. The models can serve for perturbative analysis of confined states in the systems with generalized form of electrostatic dots. This way, they can provide insight into a wider class of settings where analytical treatment is not possible. Remark that analytical solutions for bound states with non-zero energy are quite unusual in the literature where mainly zero-modes confined by electrostatic potentials are considered \cite{portnoi4,portnoi3,lucas,Ho,Schulze2,Portnoi5}.

The electrostatic field (\ref{V(x,y)}) lacks both translational and rotational symmetry. To our best knowledge, such configuration of the electric field in graphene nanoribbons have not been discussed in the literature so far.  We believe that the presented results improve our understanding of electronic properties of Dirac fermions in graphene and arm-chair nanoribbons in particular. We also think that our analysis, based on reflection and periodic symmetries, can inspire further investigation of analytically solvable models of Dirac fermions in graphene nanoribbons with  electrostatic field. We plan to report on our progress along this line in the near future.

\section*{Acknowledgement}
V.J. was supported by GA\v CR grant no.19-07117S. Partial financial support is acknowledged to Ministry of Science of Spain: PID2020-113406GB-I00 and to Junta de Castilla y Le\'on: BU229P18. V.J. and \c{S}.K. would like to thank Departamento de F\'i{}sica Te\'orica, At\'omica y \'Optica, Universidad de Valladolid, for warm hospitality.

\section*{Appendix}

Let us have a system described by the Hamiltonian $H=H_0+V(x,y)$ where $V(x,y)$ can be arbitrary Hermitian matrix with position-dependent entries. 
We set the following boundary condition in $x=0$ and $x=L$,
\begin{equation}\label{conditions2}
\Psi(0,y)=M_1\Psi(0,y),\quad \Psi(L,y)=M_2\Psi(L,y).
\end{equation}
We suppose that 
\begin{equation}\label{appendixcom}
[P_0M_1,H_0]=0,\quad [P_0M_1,V]=0,
\end{equation}
where $P_0$ is reflection at $x=0$, $P_0f(x,y)=f(-x,y)$.   
The first commutator in (\ref{appendixcom}) implies that the matrix $M_1$ has the following form
\begin{equation}\label{M1}
M_1=\boldsymbol{\nu}.\boldsymbol{\sigma}\otimes\sigma_2,
\end{equation}
where the real vector $\boldsymbol{\nu}$ is arbitrary. Therefore, there holds $[P_0M_1,H_0]=0$ both for armchair-like boundaries specified by (\ref{generalM}) and  for the infinite-mass boundary conditions given by (\ref{zig-zag-like M}) with $\nu=\pi/2$.
We suppose that the potential $V(x,y)$ is such that the second commutator in (\ref{appendixcom}) is vanishing as well. This is particularly the case for the electrostatic interaction $V(x,y)=v(x,y)\tau_0\otimes\sigma_0$ that is even in $x$. 

Let us take a generic solution $F=F(x,y)$ of $(H-E)F(x,y)=0$. We define the function $\Psi$,
\begin{equation}\label{psi0b}
\Psi(x,y)=(1+P_0M_1)F(x,y),
\end{equation}
that is an eigenstate of $H$. It is straightforward to see that the function $\Psi$ satisfies the boundary conditions in $x=0$ by construction, $\Psi(0,y)=M_1\Psi(0,y).$ The formula (\ref{psi0b}) represents a simple way how to construct spinors that follow the boundary condition at $x=0$ from a generic solution $F(x,y)$ of the stationary equation $(H-E)F(x,y)=0$. 

We require $\Psi$ to satisfy the bounday conditions at $x=L$. Substituting (\ref{psi0b}) into the second relation in (\ref{conditions2}), the corresponding equation can be brought into the following form,
\begin{equation}\label{F1b}
(1-M_2)(F(L,y)-M_2M_1F(-L,y))=0.\end{equation} 
Therefore, when $F(x,y)$ satisfies 
\begin{equation}\label{F(L,y)b}
F(L,y)=M_2M_1F(-L,y),
\end{equation}
the wave function $\Psi$ in (\ref{psi0b}) satisfies the required boundary conditions (\ref{conditions2}).


\begin{thebibliography}{99}

\bibitem{portnoi1}R. R. Hartmann, N. J. Robinson, and M. E. Portnoi, Smooth electron waveguides in graphene, Phys. Rev. B {\bf81}, 245431 (2010).

\bibitem{portnoi1b}R. R. Hartmann, Portnoi M. E., Two-dimensional Dirac particles in a P\"oschl-Teller waveguide, Sci. Rep. {\bf7}, 11599  (2017).

\bibitem{portnoi2}	D. A. Stone and C. A. Downing, M. E. Portnoi, Searching for confined modes in graphene channels: The variable phase method, Phys. Rev. B {\bf86}, 075464 (2012).

\bibitem{schulze} A. Schulze-Halberg, A. M. Ishkhanyan,  Darboux partners of Heun-class potentials for the two-dimensional massless Dirac equation, Ann. Phys. {\bf 421},  168273 (2020).

\bibitem{portnoi3}C. A. Downing, D. A. Stone, and M. E. Portnoi, Zero-energy states in graphene quantum dots and rings, Phys. Rev. B {\bf84}, 155437 (2011).

\bibitem{portnoi4} M. E. Portnoi, C. A. Downing, A. R. Pearce, and R. J. Churchill, Optimal traps in graphene,	Phys. Rev. B {\bf92}, 165401 (2015).

\bibitem{lucas} S. Kuru, J. Negro, L. M. Nieto, L. Sourrouille, Massive and massless two-dimensional Dirac particles in electric quantum dots, Preprint arXiv:2104.06676.

\bibitem{zhou}B. Zhou, B. Zhou, W. Liao, G. Zhou, Electronic transport for armchair graphene nanoribbons with a potential barrier, Phys. Lett. A {\bf374}, 761–764 (2010).

\bibitem{nascimento}  A. C. S. Nascimento, R. P. A. Lima, M. L. Lyra,   J. R. F. Lima, Electronic transport on graphene armchair-edge nanoribbons with Fermi velocity and potential barriers, Phys. Lett. A {\bf 383}, 2416-2423 (2019).






\bibitem{Akhmerov}A. R. Akhmerov, C. W. J. Beenakker, Boundary conditions for Dirac fermions on a terminated  honeycomb lattice, Phys. Rev. B {\bf 77}, 085423 (2008)

\bibitem{Brey}L. Brey, H. A. Fertig, Electronic states of graphene nanoribbons studied with the Dirac equation, Phys. Rev. B {\bf73}, 235411 (2006).

\bibitem{siegl}P. Freitas, P. Siegl, Spectra of graphene nanoribbons with armchair and zigzag boundary conditions,
Rev. Math. Phys. {\bf26}, 1450018 (2014).

\bibitem{Falco}E. McCann, V. I. Fal'ko, Symmetry of boundary conditions of the Dirac equation for electrons in carbon nanotubes, J. Phys.: Condens. Matt. {\bf 16}, 2371-2379 (2004).

\bibitem{experimentalAGNR} H. Huang, D. Wei, J. Sun, S. L. Wong, Y. P. Feng, A. H. Castro Neto, A. T. S. Wee, Spatially resolved electronic structures of atomically precise armchair graphene nanoribbons, Sci. Rep. {\bf 2}, 983
(2012).

\bibitem{NovoselovGuinea}A. H. Castro Neto, F. Guinea, N. M. R. Peres, K. S. Novoselov, A. K. Geim, The electronic properties of graphene, Rev. Mod. Phys. {\bf81}, 109-162 (2009).

\bibitem{vanOstaay} J. A. M. van Ostaay, A. R. Akhmerov, C. W. J. Beenakker, M. Wimmer, Dirac boundary condition at the reconstructed zigzag edge of graphene, Phys. Rev. B {\bf 84}, 195434 (2011).

\bibitem{siegl2}J.-C. Cuenin, P. Siegl, Eigenvalues of one-dimensional non-self-adjoint Dirac operators and applications,
Lett. Math. Phys. {\bf108}, 1757-1778 (2018).

\bibitem{superKlein}A. Contreras-Astorga, F. Correa, and V. Jakubsk\'y, Super-Klein tunneling of Dirac fermions through electrostatic gratings in graphene,  Phys. Rev. B {\bf 102}, 115429 (2020).

\bibitem{Ho} C.-L. Ho, P. Roy, On zero energy states in graphene, EPL {\bf108}, 20004 (2014).

\bibitem{Schulze2}A. Schulze-Halberg, P. Roy,  Construction of zero-energy states in graphene through the supersymmetry formalism, J. Phys. A {\bf50}, 365205 (2017).

\bibitem{Portnoi5}C. A. Downing, M. E. Portnoi, Zero-Energy Vortices in Dirac Materials, Phys. Stat. Sol. {\bf256}, 1800584 (2019).









	

		
		

		
	

	
	
	
\end{thebibliography}
\end{document}